\documentclass[aps,prd,reprint,showkeywords,superscriptaddress,showpacs]{revtex4-1}
\usepackage{bbm,amsmath,amsfonts,amssymb,indentfirst,syntonly,graphicx}
\usepackage{amsmath,amssymb,amsthm}
\usepackage{latexsym,graphicx,bbm}
\newcommand{\beq}{\begin{equation}}
\newcommand{\eeq}{\end{equation}}
\newcommand{\p}{\partial}

\begin{document}

\title{Anomalous spin of the Chern-Simons-Georgi-Glashow model}
\date{\today}
\author{Qiu-Hong Huo}
\affiliation{Laboratory of Thin Film Materials, College of Materials Science and  Engineering,
 Beijing University of Technology, 100124 Beijing , China}
 \author{Yunguo Jiang}
 \email[]{jiangyg@ihep.ac.cn}
 \affiliation{Institute of High Energy Physics, Chinese Academy of Sciences, 100049 Beijing, China}
 \affiliation{Theoretical Physics Center for Science Facilities, Chinese Academy of Sciences, 100049 Beijing,China}

\begin{abstract}
With the Coulomb gauge, the Chern-Simons-Georgi-Glashow (CSGG) model is quantized in the Dirac formalism for the constrained system. Combining the Gauss law and Coulomb gauge consistency condition, the difference between the Schwinger angular momentum and canonical angular momentum of the system is found to be an anomalous spin. The reason for this result lies in that the Schwinger energy momentum tensor and the canonical one have different symmetry properties in presence of the Chern-Simons term.
\end{abstract}
\maketitle
\section{Introduction}

Chern-Simons (C-S) field has applications in many branches of physics. It is a topological term which does not contribute to the dynamics of the gauge field, but it gives a topological mass to the gauge field \cite{Jackiw1981,Deser}. It only exist in the odd dimension of the space time, and can lead to fractional spin and fractional charge \cite{Wilczek1983,wu1984}. In the Maxwell-Chern-Simons-Higgs (MCSH) theories, both Abrikosov-Nielsen-Olesen (ANO) vortices and C-S vortices can be constructed \cite{lee1990,lee1992}. C-S vortices have ring-shaped magnetic flux, different from that of the ANO vortices with a Gauss-like shape \cite{Gudnason:2009ut}. Many C-S models are used to explain the quantum Hall effects and high temperature superconductors \cite{Girvin, Polyakov}, since C-S vortices carry fractional spin behaves as anyon-like objects.

The fractional spin in many C-S theories  has structural similarity, this is a remarkable feature. When the theory involves the Maxwell term, or the non-Abelian Yang-Mill piece, the calculation of the spin becomes subtle. A novel method to calculate the fractional spin was invented to overcome these difficulties \cite{banerjee1994,banerjee,banerjee1999}, by calculating the difference between the Schwinger angular momentum and the canonical angular momentum. The  difference is a compact form and can be interpreted as the spin of vortex. When the asymptotic form of the gauge form in a vortex configuration is used, the difference becomes the common fractional spin. A question arises naturally: why this method succeeds? In this letter, this method will be used to calculate the anomalous spin of the CSGG model. Since the scalar field in the  Chern-Simons-Georgi-Glashow  (CSGG) model is in the adjoint representation, no vortex configurations can be constructed. Combining the Gauss law and Coulomb gauge consistency condition, the solution of the gauge field is given.  With this solution, the anomalous spin term is still obtained, and this method is valid in the presence of the C-S term. The reason lies in that the Schwinger energy momentum tensor has different symmetry properties from the canonical one. In Sec. \ref{sec:spin}, we are going to explain this with more details.

The advantage of the Coulomb gauge is that there are no time derivative term in the gauge fixed action. The infrared divergence in the Maxwell-Chern-Simons (MCS) theory occurs when the Coulomb gauge is applied \cite{Jackiw1981}. There are also some ambiguities in the Yang-Mills Feynman integrals due to the absence of the time derivatives in the action \cite{Jackiw1990}. The consistency of the non-Abelian C-S theory in the Coulomb gauge at any perturbation order was investigated by Ferrai and Lazzizzera \cite{Ferrari1997}. With the pure non-Abelian C-S term, the Hamilton is zero because the C-S term contributes nothing to the dynamics. However, taking account of the Gauss law and the Coulomb gauge, the communication relations between the gauge fields vanish identically at any perturbative order \cite{Ferrari1997}. In Sec.\ref{sec:quant}, we exploit the quantization of the non-Abelian C-S theory with matter field, i.e. the CSGG model, by the Dirac quantization formalism. We will show that even with the matter field, the combination of Gauss law and the Coulomb gauge also leads to the vanish of the communication relations between the gauge fields. This is helpful to study the perturbation at any order, for example the quantum scattering amplitudes.

\section{Canonical quantization \label{sec:quant}}

The 2+1 dimensional Lagrangian of the CSGG model in the component form is written as \cite{Tekin}
\begin{align} {\cal L}=&-\frac{1}{4} F^{a}_{\mu \nu}F^{ \mu \nu,a}+ \frac{\kappa}{4}\epsilon^{\mu \nu \rho}[F^{a}_{\mu \nu}A^a_{\rho}-\frac{g}{3} f^{abc}A^a_{\mu}A^b_{\nu}A^c_{\rho}] \nonumber\\
      &+ \frac{1}{2}(D_{\mu} \phi^a)(D^{\mu} \phi^a)-\frac{m^2}{2}\phi^a \phi^a -\frac{\lambda}{4}(\phi^a \phi^a)^2,
      \label{eq:lagran}
\end{align}
where the $a,b,c$ denote the group indices, the Greek indices $\mu,\nu,\rho,\ldots=0,1,2$ denote the space-time. Under the gauge transformation, the variation of the Lagrangian has a variation term proportional to $\kappa$. Thus,  $\kappa=g^2 n/(4\pi)$ ($n\in \mathbb{Z}$) must be a quantized constant in order to leave the quantum amplitude gauge invariant. The  $\phi^a$ carries the index of gauge group, so it belongs to the adjoint representation. This gauge group is not fixed in the present discussion, it can be $SU(N)$,$SO(N)$, $USp(2N)$, etc.
If we take it to be $SO(3)$,  the well-know 't Hooft-Polyakov monopole solution exists in this model. $f^{abc}$ is an antisymmetric tensor.
$g$ and $\lambda$ are the coupling constants for $A_{\mu}$ and $\phi^a$, respectively. The gauge field strength and the covariant derivative $D_{\mu}$ are written as
\begin{align}
F^a_{\mu \nu} =&\p_{\mu}A^a_{\nu}-\p_{\nu}A^a_{\mu}+g f^{abc} A^{b}_{\mu} A^c_{\nu}, \\
 D_{\mu} \phi^a =& \partial_{\mu}\phi^a+ g f^{abc}A^b_{\mu} \phi^c.
 \end{align}
The Euler-Lagrangian equation can be calculated, which are
\begin{align}
D_{\nu}F^{\nu \mu,a} + \frac{\kappa}{2} \epsilon^{\mu \nu \rho} F^a_{\nu \rho}+ g f^{abc} D^{\mu}\phi^b \phi^c=&0, \label{eq:el1}\\
\left[D_{\nu} D^{\nu}-m^2 +\lambda (\phi^b \phi^b)\right]\phi^a=&0.
\end{align}
The canonical momentum of the field is defined to be $\pi_X \equiv \partial {\cal L}/\partial \dot{X} $, where $X$ denotes any field. With this definition, the canonical momentum of the gauge and the matter fields are
\begin{align}
\pi^{\mu,a}&=-F^{0\mu,a}+\frac{\kappa}{2} \epsilon^{0\mu \rho} A^{a}_{\rho}, \\
\pi^a_{\phi}&=D^0 \phi^a. \end{align}
For $\mu=0$, $\pi^{0,a}=0$ gives a constraint. For $\mu=i$, one has $\pi^{i,a}=-F^{0i,a}+\frac{\kappa}{2} \epsilon^{0ij} A^{a}_{j},$, which contains a time derivative term. Therefore, the gauge fields $A^a_{\mu}$ has dynamics in the CSGG model.
The canonical Hamiltonian is written as
\begin{align}
{\cal H}_C=& - \frac{1}{2} \pi^a_i \pi^{i,a} -A^a_0 \p^i \pi^a_i-gf^{abc}\pi^a_i A^b_0 A^{i,c}+\frac{\kappa}{2}\epsilon^{ij}\pi^a_i A^a_j\nonumber\\
&+\frac{1}{2}\pi^a_{\phi}\pi^a_{\phi}-gf^{abc}\pi^a_{\phi}A^b_0 \phi^c -\frac{\kappa^2}{8}A^{i,a}A^a_i+\frac{1}{4}F^a_{ij}F^{ij,a} \nonumber \\
&+\frac{\kappa g}{12} \epsilon^{\mu \nu \rho} f^{abc}A^a_{\mu}A^b_{\nu}A^c_{\rho} -\frac{\kappa}{4}\epsilon^{ij}A^a_0 F^a_{ij} \nonumber\\
&-\frac{1}{2}D_i\phi^a D^i \phi^a +\frac{m^2}{2} \phi^a \phi^a + \frac{\lambda}{4} (\phi^a \phi^a)^2.
\end{align}
The definition of $\pi^a_{\mu}$ gives that
\beq \Lambda^a_1=\pi^a_0 \approx 0, \eeq
which is called the primary constraint by Dirac \cite{Dirac}. The total Hamiltonian is given by
\beq {\cal H}_{T}={\cal H}_C + \eta^a\Lambda^a_1, \eeq
where $\eta^a$ is the Lagrangian multiplier. The consistency condition requires that
\begin{align}
 \dot{\Lambda}^a_1 =&\{\pi^{0,a}, \int d^2x {\cal H}_T \} \nonumber \\
  =&\p^i\pi^a_i+\frac{\kappa}{2} \epsilon^{ij} \p_i A^a_j+gf^{abc}A^{i,b}\pi^c_i +gf^{abc}\pi^b_{\phi}\phi^c \nonumber \\
   \equiv&\Lambda^a_2 \approx 0.\label{eq:lam2}
  \end{align}
This new constraint $\Lambda^a_2$ is the secondary constraint, which is the Gauss law. No further constraint is produced when $\dot{\Lambda}^a_2 $ is considered. One can verify that $\{ \Lambda^a_1, \Lambda^a_2 \}\approx0$, this means both $\Lambda^a_1$ and $ \Lambda^a_2$ are first class constraints. Dirac conjectured that all the first class constraints can be added to the Hamiltonian and the dynamics of the system does not change \cite{Dirac}. Costa et al. proved that Dirac conjecture is valid when all the constraints are first class \cite{costa1985}. $\pi^a_0$ is not the physical degree of freedom, since photon has only two physical degree. Thus, $\Lambda^a_1$ can be eliminated without changing the dynamics, i.e., the Hamiltonian equations. The constraint $\Lambda^a_2 $ is the generator of the gauge transformation \cite{wang2009}.

In order to quantize the system, two gauge fixing conditions should be introduced in order to fix the gauge. First we consider the Coulomb gauge
\beq \Omega^b_1\equiv \p_i A^{i,b} \approx 0. \eeq
This gauge has advantages when we consider the static configuration of the soliton system, since it has no time derivative. In Dirac's procedure, the extended Hamiltonian can be obtained by adding the secondary first class constraint, which is written as
\begin{align} {\cal H}_{E}=&{\cal H}_C + \eta^a_1\Lambda^a_1+\eta^a_2 \Lambda^a_2 \nonumber \\
    \cong &H_T. \end{align}
The sign "$\cong$" means that the Lagrangian multiplier $\eta^a_2$ can be absorbed to $A^a_0$ by redefining $A^a_0 {'} =A^a_0+\eta^a_2$. This is an evident support for the Dirac conjecture.
Another condition $\Omega^b_2$ should be consistent with $\Omega^b_1$, Requiring $\{\Omega^b_1, H_T\}\approx 0$, one obtains
\begin{align} \label{eq:ome2}
 \Omega^b_2=&\{\Omega^b_1, H_T \}  \nonumber\\
 =&-\p_i\pi^{i,a}+\frac{\kappa}{2} \epsilon^{ij} \p_i A^a_j + \p^i D_i A^a_0.
\end{align}
The non-trivial communication relations are listed as
\begin{align} \{ \Lambda^a_1(x), \Omega^b_2(y)\} &= -\p^i_yD^{ab}_i(y) \delta(x-y),  \\
\qquad \{ \Lambda^a_2(x), \Omega^b_1(y)\} &=- \p^i_y D^{ab}_i(x) \delta(x-y),  \end{align}
where $D^{ab}_i\equiv \p_i \delta^{ab} + g f^{abc} A^c_i $. Now $\Lambda^a_i, \Omega^b_j$ can be considered to be
the secondary class constraints, which form a non-singular matrix $C^{ab}(x,y)$
\beq C^{ab}(x,y)=\left(
                   \begin{array}{cc}
                     0 &  -\p^i_yD^{ab}_i(y)\\
                     - \p^i_y D^{ab}_i(x) & 0 \\
                   \end{array}
                 \right)\delta(x-y). \eeq
The inverse matrix of the constraint $C^{ab}(x,y)$ can be solved by introducing a Green function $G^{cb}(x,y)$ \cite{Schwinger1962,Ferrari1997},
\beq D_i^{ac}(x)\p^i_x G^{cb}(x,y)=\delta^{ab}\delta(x-y).\eeq
Supposing $G^{ab}(x,y)$ has a good behavior at infinity, one obtains
\beq (C^{-1})^{ab}(x,y)=\left(
                   \begin{array}{cc}
                     0 & G^{ab}(x,y)\\
                     - G^{ab}(x,y) & 0 \\
                   \end{array}
                 \right). \eeq
The quantized communication relation (QCR) should be realized by the Dirac bracket, which is defined as
\begin{align}
\{F^a(x),D^b(y) \}^*=&\{F^a(x),D^b(y) \}-\int d z d w \nonumber \\
 &  \{F^a(x),\Lambda^c_i(z) \} \cdot(C^{-1})^{cd}_{ij}(z,w)\nonumber \\
 & \cdot \{\Omega^d_j(w),D^b(y) \}, \end{align}
where $C^{ab}_{ij}(x,y)\equiv \{\Lambda^a_i(x), \Omega^b_j(y)\}$.
To quantize the theory, we need to replace the Dirac brackets with the commutators. The QCR of the gauge field $A^a_{i}$ and $\pi^b_j$ is calculated to be
\beq
[A^a_i(x), \pi^b_j(y) ]= \delta_{ij} \delta^{ab} \delta(x-y).
\eeq
One can also verify that $[A^a_i(x), A^b_j(y)] = 0 $. This gives the correct quantization result for the CSGG model.
In a pure non-Abelian C-S theory, the C-S field theories are shown to be finite and free in the Coulomb gauge \cite{Ferrari1997}. It was questioned that whether such commutation relation (CR) vanish at all perturbative orders in presence of the matter field. Our analysis above show that the CR of the gauge fields vanish identically  by taking into account the Gauss law and the Coulomb gauge fixing in the CSGG model.

\section{Anomalous spin \label{sec:spin}}

Banerjee and Mukherjee proposed a novel method to calculate the fractional spin term in C-S  system \cite{banerjee}. This spin is interpreted as the difference between the angular momentum obtained by modifying Schwinger's energy-momentum tensor with Gauss constraint, and the canonical (Noether) angular momentum. Also the fractional spin term can also be calculated by the canonical angular momentum in the quantized system \cite{wang2010,Huang:2008zzb}. Here, we take use of this method to calculate the spin of CSGG model, and obtain an anomalous spin term. The success of this method is that the  symmetric properties of the modified Schwinger's energy momentum tensor is different with that of the canonical energy momentum.

The Schwinger's energy-momentum tensor is given by \cite{schwinger1962}
\begin{align} \Theta_{\mu \nu}= &\frac{2}{\sqrt{-g}}\frac{\delta S}{\delta g^{\mu \nu}} \nonumber \\
    =& F^a_{\mu \rho} F^{\rho,a}_{\nu}+\frac{1}{2}\left( D_{\mu} \phi^a D_{\nu} \phi^a +D_{\nu} \phi^a D_{\mu}\phi^a \right) -g_{\mu \nu}{\cal L}.
  \label{eq:theta1} \end{align}
It is evident that this tensor is symmetric, i.e., $T^{\mu \nu}=T^{\nu \mu}$. The is because the metric of the system is symmetric, $g^{\mu \nu}=g^{\nu \mu}$. However, the C-S term does not contain the metric. Thus, the C-S term has no contribution to the Schwinger energy momentum tensor.  In present the constraint, a more general expression for $\Theta_{\mu \nu}$ can be given by
\beq  \Theta^T_{\mu \nu}= \Theta_{\mu \nu}+ \Gamma^a_{\mu \nu }G^a,  \label{eq:theta2}\eeq
where $\Gamma^a_{\mu \nu}$ is the Lagrangian multiplier. Since $G^a$ is the generator of gauge transformations, $ \Theta^T_{\mu \nu}$ is gauge invariant on the constraint surface \cite{banerjee}, i.e.,
\beq \{  \Theta^T_{\mu \nu}, G^a\} \approx 0. \eeq
In order to keep the correct spatial translation , a suitable choice for $\Gamma^a_{0i}$ is that $\Gamma^a_{0i}=-A^a_i$ \cite{banerjee}. Thus, the Schwinger energy momentum conserves both the gauge and Lorentz symmetry.

The angular momentum operator is defined as
\beq L=\int d^2 x \epsilon^{0ij}x_i  \Theta^T_{0j}. \label{angl}\eeq
Considering Eq.(\ref{eq:theta1}) and Eq.(\ref{eq:theta2}), $\Theta^T_{0j}$ is calculated to be
\beq \Theta^T_{0j} \approx \pi^a_k F^{k,a}_j-\frac{\kappa}{2}\epsilon_{0kl}A^{l,a}F^{k,a}_j+D_j\phi^a \pi^a_{\phi}. \label{tensor}\eeq
Here, we use the Dirac weak equality condition.
Substituting Eq.(\ref{tensor}) into Eq.(\ref{angl}), one obtains that
\begin{align} \label{eq:schang}
L_S=&\int d^2x \epsilon^{ij}x_i \Theta^T_{0j} \nonumber \\
  = &\int d^2 x \,\epsilon^{ij}(x_i\pi_k F^{k,a}_j+x_iD_j \phi^a \pi^a_{\phi}) \nonumber \\
   &+\frac{\kappa}{2} \int d^2x \,\,x_i A^{j,a}F^{i,a}_j. \end{align}
Notice that only the canonical variables are used.

The canonical energy-momentum tensor is defined  as \cite{banerjee}
\beq T_{\mu \nu}=\frac{\p {\cal L}}{\p (\p^{\mu} \phi^a)}\p_{\nu}\phi^a+\frac{\p {\cal L}}{\p(\p^{\mu} A^a_{\rho})}\p_{\nu}A^a_{\rho}-g_{\mu\nu} {\cal L},\eeq
and $T_{0i}$ is calculated to be
\beq T_{0i}\approx \pi^a_{\phi}\p_i\phi^a+\pi^a_{k}\p_i A^a_{k}, \eeq
where the term proportional to primary constraint $\Lambda^a_0$ is ignored. Notice that, $T_{0i} \neq T_{i0}$, since $T_{i0}$ has a time derivative term $\pi^a_{i}\p_0 A^a_{i}$, which should be converted into the canonical variables.
The asymmetry of the canonical tensor originates from the C-S term.
The canonical angular momentum tensor is defined as
\beq M_{\mu \nu}=\int d^2 x \, \left[x_{\mu}T_{0\nu}-x_{\nu}T_{0\mu}+\pi^a_{\rho} \Sigma^{\rho \lambda}_{\mu\nu} A^a_{\lambda}\right].\eeq
For the scalar field, $\Sigma^{\rho \lambda}_{\mu\nu}=0$,  while for the vector field, $\Sigma^{\rho \lambda}_{\mu\nu}=\delta^{\rho}_{\mu} \delta^{\lambda}_{\nu}-\delta^{\rho}_{\nu} \delta^{\lambda}_{\mu}$. Thus, the angular momentum $L$ can be written as
\begin{align} L_C=&\frac{1}{2}\epsilon^{ij}M_{ij} \nonumber \\
=&\epsilon^{ij}\int d^2 x\,\,\left[ x_i \pi^a_{\phi} \p_j \phi^a + x_i \pi^a_k \p_j A^a_k +\pi^a_i A^a_j\right]. \label{eq:canang} \end{align}
Comparing the Schwinger angular momentum $L_S$ in Eq.(\ref{eq:schang}) and the canonical angular momentum $L_C$ in Eq.(\ref{eq:canang}), it seems that $L_S$ contains the C-S contribution while $L_C$ does not. This is an illusion because the definition of $\pi^a_i$ includes the C-S related term. In Eq.(\ref{eq:schang}), the last term $\frac{\kappa}{2} \int d^2x \,\,x_i A^{j,a}F^{i,a}_j$ diminishes with the C-S contribution in $\pi^a_i$. Thus, $L_S$ is equivalent to the angular momentum in the Maxwell-Georgi-Glashow (MGG) theory.

The difference between the Schwinger angular momentum and the canonical angular momentum is written as
\beq K=L_S - L_C \approx -\int d^2 x \, \p^k \left[ \pi^a_k \epsilon^{ij}x_i A^a_j \right], \eeq
which is a total boundary term. In the process of calculation, the constraint  $\Lambda^a_2$ is considered. The matter component $\phi^a$ contributes the same gradient to $L_S$ and $L_C$, and then disappears in $K$. For singular configurations, i.e., in the presence of C-S vortices, $K$ does not vanish \cite{banerjee}. In the following, we will show that $K$ is an anomalous spin term even without the existence of singular configurations.


Combining the secondary constraint $\Lambda^a_2$ (Gauss law) in Eq.(\ref{eq:lam2}) and the gauge fixing condition $\Omega^a_2$ in Eq.(\ref{eq:ome2}), one obtains
\beq \label{eq:aai} \kappa \epsilon^{ij} \p_i A^a_j=-\p_iD^i A^a_0+g f^{abc} (\pi^b_{\phi}\phi^c+\pi^{i,b}A^c_i)\equiv J^a. \eeq
Integrating over the 2-dimensional spatial surface on both sides of the Eq.(\ref{eq:aai}), one obtains the non-Abelian flux of the system,
\beq \label{eq:flux} \Phi^a \equiv \int \, d^2x \, \epsilon^{ij} \p_i A^a_j=\frac{Q^a}{\kappa},
\eeq
where $Q^a= \int \, d^2x J^a$. Eq.(\ref{eq:aai}) contains no time derivative term, one can construct a solution for $A^a_i$, which is written as
\beq A^a_i=-\epsilon^{ij}\frac{x^j}{|x^2|}\frac{Q^a}{2\pi \kappa}. \label{eq:asolu}\eeq
Here the soliton configuration is not referred. Substituting Eq.(\ref{eq:asolu}) into $K$, one obtains
\beq K=-\frac{Q^a Q^a}{4\pi \kappa}.\eeq
Up to now, the analysis is general in a sense that no soliton Ansatz is accounted, no specific gauge group is specified. The formula of $K$ can be called the anomalous spin term.

In the CSGG model, no vortex solutions can be constructed, because $\phi^a$ is in the adjoint representation. Thus, whether $K$ stands for a fractional spin  term is in doubt. If there are three spatial dimension, 't Hooft-Polyakov monopole  solution really exists in this model, namely the complex monopoles \cite{Tekin}. Since we work in only 2 spatial dimensions, the topological excitations forbid the existence of monopoles. In order to keep the gauge invariance, $\kappa$ should be quantized as mentioned before, otherwise, the action changes under gauge transformation .  For $\kappa=g^2 n/(4\pi)$, one has $K=-Q^aQ^a/(g^2 n)$. One can interpret such $K$ as the fractional spin, it can be arbitrary smaller if $n \to \infty$.  Our analysis does not depend on the matter components, thus it holds also for the vortex system. For instance, if one replaces the adjoint scalar $\phi^a$ with a fundamental complex field $\varphi$, one can construct a non-Abelian C-S vortices \cite{Gudnason:2009ut}. In the vortex configuration, the asymptotic behavior of $D_{i} \varphi$ leads to $Q^aQ^a=m^2$. Thus, $K=-\frac{m^2}{g^2n}$, where $m$ stands for the winding number of vortices, and $n$ is an arbitrary integer.

\section{Conclusion}

In this letter, we investigate the 2+1 dimension CSGG theories with arbitrary gauge group. By the Dirac formalism for the constrained system, we obtain two constraints, which are all first class. Taking the coulomb gauge and its consistency as gauge fixing conditions, the Dirac brackets of the gauge fields are deduced. Replacing the Dirac brackets with the communication relations, the system is quantized in the canonical quantization formalism. Schwinger angular momentum and Noether angular momentum of the system are calculated respectively. The difference of them is found to be a a compact form. Combining the Gauss law and Coulomb gauge consistency condition, the solution of the gauge field is given. With this solution, the difference of the two angular momentum is converted to be an anomalous spin term.

In our analysis, the reason of the difference between the Schwinger and the canonical angular momentum is given. The C-S term disappears when the action is varied with the metric,  thus the Schwinger energy momentum tensor are symmetric about the indices. Meanwhile, the C-S term do contribute to the canonical energy momentum tensor, which are asymmetric when permuting the indices. Therefore, the difference originate  from the C-S term in the Lagrangian. The matter fields and the Maxwell term contribute the same for both angular momentum, thus they do not appear in $K$. The anomalous spin term is also found in the canonical angular momentum of C-S system at quantum level \cite{Huang:2008zzb,wang2010}, which support the analysis here.

Previously, the fractional spin of C-S systems is realized  with the vortex configurations \cite{lee1990,lee1992,Wilczek1983,wu1984,Jackiw1981,banerjee}. In this letter, no vortex configuration is used to deduce $K$. The Gauss law together with the Coulomb gauge are sufficient to guarantee the existence of the anomalous spin. Thus, the anomalous spin exist no matter whether there are soliton solutions or not. C-S vortices  have broken axial symmetry, multi-fractional vortex-centers, ring-like flux structure, etc \cite{Gudnason:2009ut}. When the Ansatz of the vortex configuration is considered, the conserved charge $Q^a$ is related to the winding number of the vortices. In this way, $K$ can be explained as the fractional spin. In 3+1 dimension, C-S term violates the both the Lorentz and parity invariance \cite{carroll1990}, but not violates the gauge invariance up to a total derivative term. Thereby, the Lorentz symmetry holds for $L_S$ but is broken in $L_C$ in 3+1 dimension. In 3+1 dimension, also 't Hooft-Polyakov monopole configuration exist in the CSGG model. It is an interesting future work  to take use of the explicit Ansatz of monopoles to calculate $K$, which may be related to the winding number of monopoles.

\begin{acknowledgments}
 The work of Q.~H.~Huo was financially supported by the National Natural Science Foundation of China (NSFC) (Grant Nos. 51032002, 11074017 and 11174021), the IHLB (Grant No. PHR 201007101), the Beijing Nova Program (Grant No. 2008B10), the Beijing Natural Science Foundation (Grant No 1102006), and the Basic Research Foundation of Beijing University of Technology. The work of Y.~G.~Jiang was funded by the National Natural Science Fund of China under Grant Nos. 10875129, 11075166.
\end{acknowledgments}

\end{document}